# PROP - PATRONAGE OF PHP WEB APPLICATIONS

Sireesha C[1], Jyostna G[2], Raghu Varan P[3] and P R L Eswari[4]

Centre for Development of Advanced Computing, Hyderabad, India

## ABSTRACT

*PHP is one of the most commonly used languages to develop web sites because of its simplicity, easy to learn and it can be easily embedded with any of the databases. A web developer with his basic knowledge developing an application without practising secure guidelines, improper validation of user inputs leads to various source code vulnerabilities. Logical flaws while designing, implementing and hosting the web application causes work flow deviation attacks. In this paper, we are analyzing the complete behaviour of a web application through static and dynamic analysis methodologies.*

## 1. INTRODUCTION

Internet and web have made the entire world come together. Also now-a-days web is being used heavily to offer citizen services including banking and governance services. However, these advances in Internet technologies are being exploited to cause adverse effects. Vulnerabilities in web applications are used as vehicles to launch various attacks. According to Symantec survey report-2013 [1], small businesses and organizations are being targeted by attackers. Popular web application threats according to OWASP [2] include SQL injection [3], Cross-Site Scripting (XSS) [4], Authentication and Authorization bypass [5][6], Session Hijacking [7], Cross-Site Request Forgery (CSRF) [8]. Most popular threat among these is SQL injection, which targets backend database through the web application. The attackers make use of the compromised web applications and acquire unauthorized access to the database. The [9] shows the risk factors of top 10 attacks incurred on specific application or organization's technical impact and business impacts.

XSS attack is launched by injecting malicious code through user supplied data. Popularly this attack is made through java script code; attacker gives malicious web link in user input field which internally calls or redirects to the web link supplied by the attacker. CSRF is another dangerous attack in which the attacker inserts malicious links in the forms or forums of a legitimate website which tempt users to click on those links leading to malicious activity.

Another attack is "Session Hijacking" where the attacker focuses more on finding the weaknesses in session implementation. By using session fixation, prediction and capture methods attacker gets session information which can be misused. The session fixation [10] occurs when an attacker is able to trick the user in using a predefined session ID of his choice. Usually the session ID is passed as URL parameter along with the requested page information from the client. If the web server fills the session details with the predefined session ID from the client without regenerating a new session ID then there is a possibility of attack. Another possible attack is sequence bypass attack, where an attacker is able to access the unauthorized pages with the same privileges or by forcible browsing.

In addition to the above, Code Injection and command Execution are other popular attacks. In Code Injection attack, malicious code is added as part of the application itself, which gets executed when application is accessed. Shell code falls under this category. In Command Execution attack, attacker injects and executes commands through vulnerable applications.

Configuration file settings are also exploited for launching the attacks. For example in PHP [11][12], if register_global is turned 'on' in configuration file it automatically takes data from the super global arrays ($_GET, $_POST, $_SERVER, $_COOKIE, $_REQUEST and $_FILE) and assigns them to global variables, means $_POST['password'] would automatically assigned to global variable $password. These global variable details are used by attacker to gain unauthorized access to the application. All these attacks are made by compromising either web application or exploiting the configuration details of .config files. In order to protect from these attacks, various research efforts are made in developing browser side as well as web application side security solutions. Through this paper we represent **PROP**- PatROnage of PHP Web Applications, which analyzes and detects the source code vulnerabilities and prevents the run time execution attacks. This security solution is implemented and tested for PHP based web application and results are promising.

## 2. EXISTING SOLUTIONS

In order to detect & prevent web application attacks, source code as well as run time analysis approaches [13] [14] [15] are used. Existing solutions pixy [16], rips [17], MIMOSA [18] and IBM Rational AppScan [19] require scripting code of web application in order to detect the vulnerabilities. Swaddler [20] is a solution, in which vulnerabilities are detected by analyzing the state of web application based on session values at PHP interpreter level during runtime. Another solution Acunetix web vulnerability scanner [21], audits web applications by checking for exploitable hacking vulnerabilities through static analysis. Nemesis [22] approach addresses the Authentication and Authorization bypass attacks with programmer-supplied access control rules on files and database entries. To provide the security at web application level another possible solution is the use of Web Application Firewalls (WAF) [23]. But WAFs are designed by white listing the rules. The rule set of the WAFs describes the behaviour of the application. But these WAFs are failing to prevent the Session Hijacking; Privilege Escalation and Logical flaws exist in web applications due to the inability in white listing the rules of defected code and session maintenance.

In this paper we propose **PROP** to detect source code vulnerabilities like XSS, SQL Injection, Code Injection, Command Injection, File Inclusion and File Manipulation attacks and to detect & prevents the work flow deviation attacks like SQL injection; authentication & authorization bypass through session stealing and sequence bypass attacks at run time execution. The solution works without disturbing the application database and without opening any external ports.

## 3. APPROACH

PROP includes Static Analyzer and Dynamic Analyzer which follows the source code analysis and run time analysis techniques respectively. Figure 1 shows the working functionality of PROP. Static Analyzer analyses the source code of PHP web application and detects the source code vulnerabilities. It maintains the vulnerability checklist which includes the identified native PHP vulnerable functions list and sources from which the vulnerabilities are exploited like user input, file and database access methods. Analysis starts by tokenizing the source code, parsing and identifying the vulnerabilities against provided vulnerable functions check list. And it generates a report on identified vulnerabilities and the report includes the information like how many files it scanned, detected vulnerabilities list and total scanning time. Detected vulnerabilities list contains file name and line number of the vulnerability. It also saves the vulnerabilities report in pdf format for further analysis.

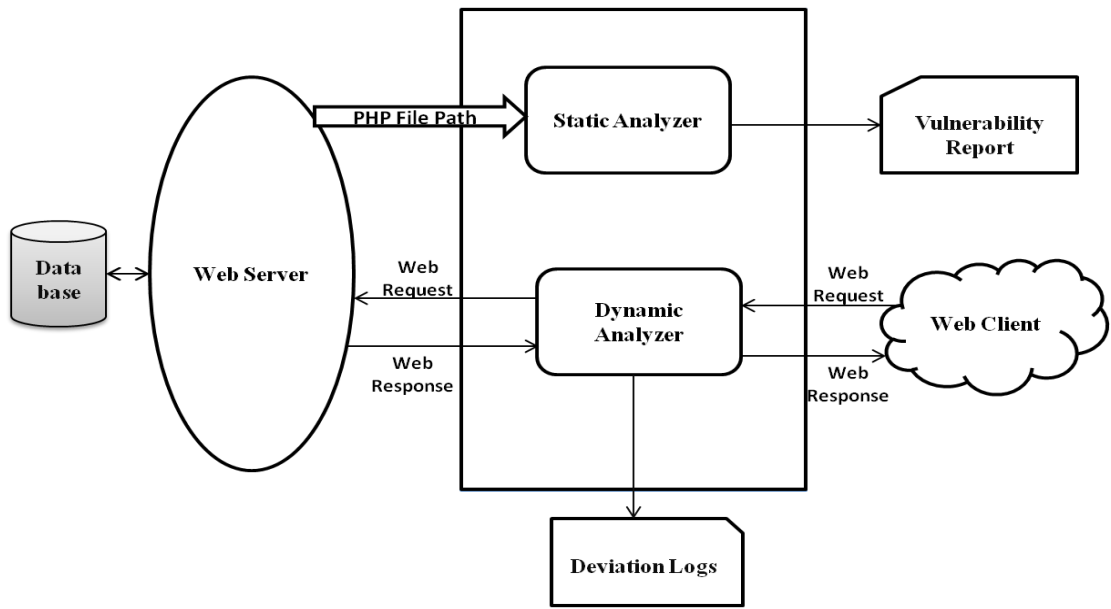

Figure 1: PROP

Dynamic Analyzer of PROP analyzes the run time execution flow by capturing the web communication. It captures web request and response messages along with the session flags. These details are collected to create a behaviour model of the web application and are stored in database at the server. Session flag in the model indicates the existence / non-existence of the session. This behaviour model is enforced at runtime along with the details like user agent and client IP address to detect the work flow deviation attacks. Figure 2 shows functionality of Dynamic Analyzer.

PROP Dynamic Analyzer is carried out in two phases: Training Phase and Runtime Enforcement Phase. During the Training Phase, PROP monitors the web application behaviour in attack free environment. It uses the spidering technique [24] to crawl internally to each and every web page and generates profiles and constructs the model by covering the complete behaviour of the web application. During Runtime Enforcement Phase, along with the web request the user agent and client IP address are also monitored and model is enforced to detect work flow deviation attacks. The detected deviations are reported for further analysis.

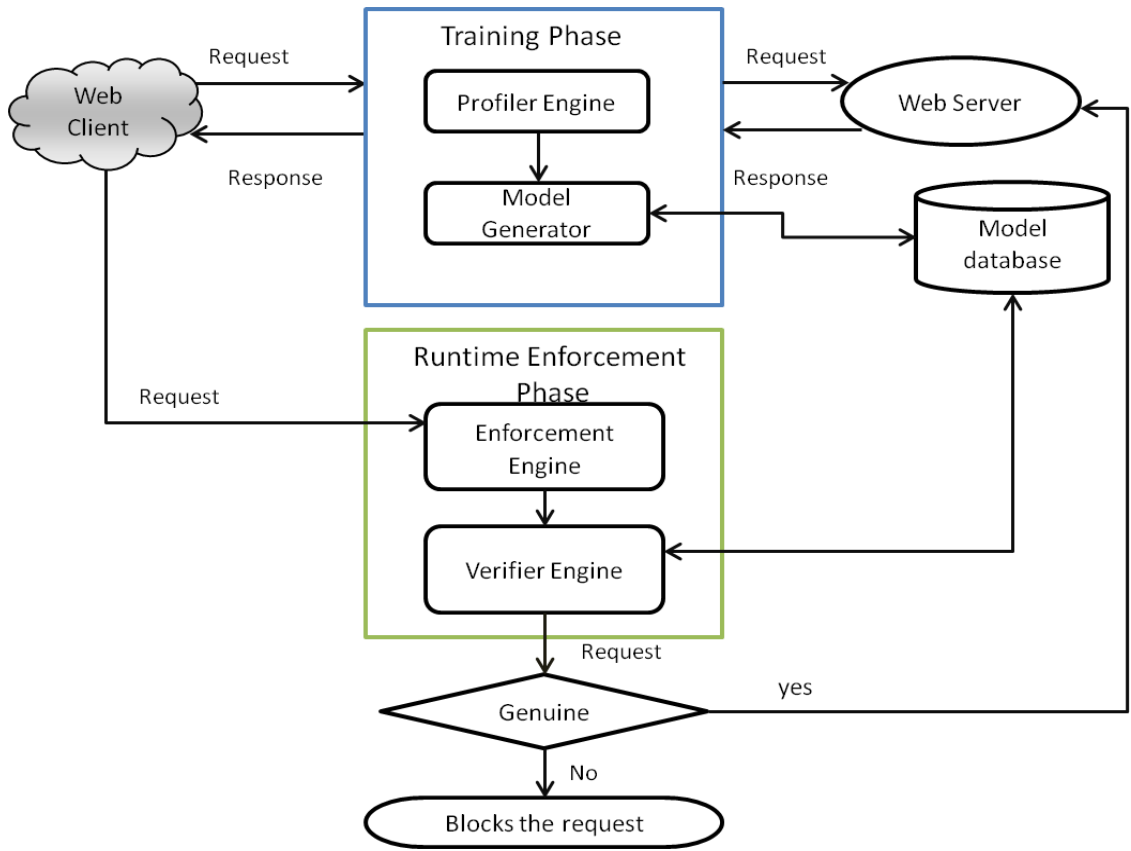

Figure 2: Functionality of Dynamic Analyzer

## 4. DESIGN LAYOUT

### 4.1. Static analyzer

PROP Static Analyzer performs 2 types of analysis: PHP Configuration file analysis and PHP source code analysis. PHP Configuration file analysis reads the native PHP configuration file (php.ini) and checks configured settings and display the mis-configured setting with current value and recommended value.

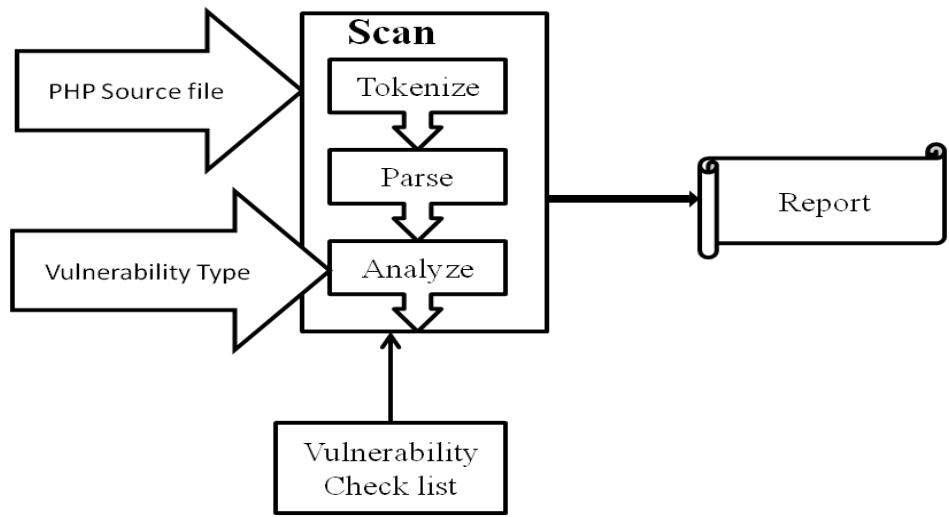

Figure 3: Functionality PHP Source code analyzer

PHP Source code analyzer analyzes the source code, it first identifies and lists the vulnerable functions in the native PHP script which causes Cross-Site Scripting, SQL Injection, Command Injection, Code Injection, File Inclusion and File Manipulation vulnerabilities and also lists secure functions to prevent these vulnerabilities from exploitation.

To analyze the PHP source code, the PHP script is split into tokens. These tokens are used for further analysis. Each token is represented in an array with token identifier, the line number and token value. And tokens are analyzed against the configured vulnerable functions list, meanwhile it creates a dependency stack, a file stack, list of declared variables and several registers to indicate whether it is currently scanning a function, or class. If any vulnerable function is detected, it creates a new parent and it checks parameters of that function by backtracking. And if any vulnerable parameter found it adds as a child to that parent. And that node the details are sent to for reporting. Figure 3 shows the functionality of PHP Source code analyzer. The same procedure is repeated for all the tokens. And finally it displays the detected vulnerabilities and can be saved in pdf format. Figure 4 shows the pdf report of Source code analyzer.

**VULNERABILITY DETAILS**

VulnerabilityNumber : 1
Vulnerability FileName : C:/xampp/htdocs/empldir_php4t/AdminMenu.php
VulnerabilityName : Cross-Site Scripting
Vulnerable Line : 114: printf printf("Debug: query = %s<br>\n", $Query_String); // db_mysql.inc

VulnerabilityNumber : 1
Vulnerability FileName : C:/xampp/htdocs/empldir_php4t/AdminMenu.php
VulnerabilityName : Cross-Site Scripting
Vulnerable Line : 131: query $db_fill->query ($sql_query);

VulnerabilityNumber : 1
Vulnerability FileName : C:/xampp/htdocs/empldir_php4t/AdminMenu.php
VulnerabilityName : Cross-Site Scripting
Vulnerable Line : 153: query $db_look->query ("SELECT " . $field_name . " FROM " . $table_name . " WHERE " . $where_condition);

Figure 4: Sample pdf report

## 4.2. Dynamic Analyzer

Working functionality of the Training Phase and Runtime Enforcement Phase are as follows.

### 4.2.1. Training Phase

Figure 5 shows the PROP Dynamic Analyzer functioning at Training Phase with **Profiler Engine** and **Model Generator** modules. **Profiler Engine** captures the web communication for different roles. For each role Profiler Engine collects the web request in the form of request header information, records and passes to the web server. The response from the web server is forwarded to the web client. Along with the request and response information, Profiler Engine also records the sequence of web requests with respect to current and previous web request state.

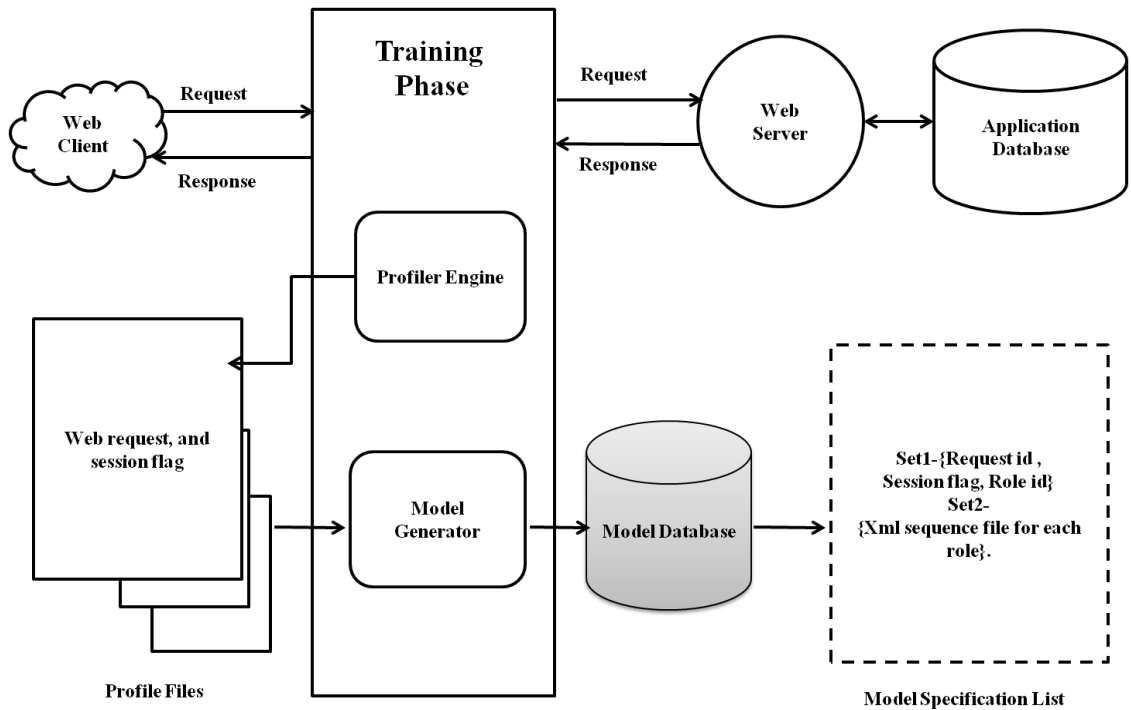

Figure 5 : PROP Dynamic Analyzer functioning at Training Phase

For each web request a separate communication id is created to differentiate between requests coming from web clients. And corresponding request header information is saved in a file with name "communicationid_request". It also extracts the cookie id from the header to check the session existence.. The session flag for that request is saved in "communication id_Srequest" file name. And also sequence of pages crawled by each role user is saved in a "roleid.xml" file. For example, with request communication id to be 1, corresponding header information is saved in 1_request and session flag is saved in 1_Srequestfile names.

After recording the request information it forwards the request to the web server for processing. The response from the web server is collected and forwarded to the web client.

Profiler Engine internally spiders each web page and collects the request and response information. Spider covers all the web pages internally for strengthening the model of the application. The same process is repeated for all the roles of the web application.

**Model generator** is another module in Training Phase which works in offline mode. It analyzes the profile records based on the communication id and role, builds a relational model database for the particular web application behaviour. It first reads the "communicationid_request" file and creates a request id based on the method of calling and requested resource name. If requested URL is http://example.com/login.php using GET method, request id becomes *GET_login.php*. From the corresponding "communicationid_Srequest" file reads the session flag. Based on the profile records it creates Model database with 2 different types of model sets.

Model set1 represents MySQL database table and each row contains communication id, request id, session flag and role. Figure 6 shows the Model set1.

```
mysql> select * from req_res_stat;
+-----+--------+------------------------------+-------------+------+
| sno | convid | reqresid                     | sessionFlag | role |
+-----+--------+------------------------------+-------------+------+
|   1 |      1 | GET_Default.php              |           0 |    0 |
|   2 |      2 | GET_login.php                |           0 |    0 |
|   3 |      3 | GET_about.php                |           0 |    0 |
|   4 |      4 | GET_installationguide.php    |           0 |    0 |
|   5 |      5 | GET_forgotpassword.php       |           0 |    0 |
|   6 |      6 | GET_login.php                |           0 |    0 |
|   7 |      7 | GET_login.php                |           0 |    0 |
|   8 |      8 | POST_login.php               |           0 |    0 |
|   9 |      9 | GET_home.php                 |           1 |    0 |
|  10 |     10 | GET_calender.js              |           1 |    0 |
|  11 |     11 | GET_home.php                 |           1 |    0 |
|  12 |     12 | GET_jquery-ui.css            |           1 |    0 |
|  13 |     13 | GET_preferences.php          |           1 |    1 |
|  14 |     14 | GET_formwizard.js            |           1 |    1 |
|  15 |     15 | GET_allreports.php           |           1 |    1 |
|  16 |     16 | GET_risk_assesement.php      |           1 |    1 |
|  17 |     17 | GET_logout.php               |           1 |    1 |
|  18 |     18 | GET_index.php                |           0 |    1 |
|  19 |     19 | GET_login.php                |           0 |    1 |
|  20 |     20 | POST_login.php               |           0 |    1 |
|  21 |     21 | GET_home.php                 |           1 |    1 |
|  22 |     22 | GET_calender.js              |           1 |    1 |
|  23 |     23 | GET_about.php                |           1 |    1 |
|  24 |     24 | GET_jquery-ui.css            |           1 |    1 |
|  25 |     25 | GET_formwizard.js            |           1 |    1 |
|  26 |     26 | GET_installationguide.php    |           1 |    1 |
|  27 |     27 | GET_forgotpassword.php       |           1 |    1 |
|  28 |     28 | GET_home.php                 |           1 |    1 |
|  29 |     29 | GET_preferences.php          |           1 |    2 |
|  30 |     30 | GET_allreports.php           |           1 |    2 |
|  31 |     31 | GET_crawl.php                |           1 |    2 |
|  32 |     32 | GET_requestscan.php          |           1 |    2 |
|  33 |     33 | GET_schedule_information.php |           1 |    2 |
|  34 |     34 | GET_risk_assesement.php      |           1 |    2 |
|  35 |     35 | GET_abcd.php                 |           1 |    2 |
|  36 |     36 | GET_abcd4m.php               |           1 |    2 |
|  37 |     37 | GET_logout.php               |           1 |    2 |
|  38 |     38 | GET_index.php                |           0 |    2 |
|  39 |     39 | GET_login.php                |           0 |    2 |
|  40 |     40 | POST_login.php               |           0 |    2 |
|  41 |     41 | GET_home.php                 |           1 |    2 |
|  42 |     43 | GET_home.php                 |           1 |    2 |
|  43 |     42 | GET_calender.js              |           1 |    2 |
|  44 |     44 | GET_jquery-ui.css            |           1 |    2 |
|  45 |     45 | GET_formwizard.js            |           1 |    3 |
|  46 |     46 | GET_preferences.php          |           1 |    3 |
|  47 |     47 | GET_usermanagement.php       |           1 |    3 |
|  48 |     48 | GET_usermanagement1.php      |           1 |    3 |
|  49 |     49 | GET_manager.php              |           1 |    3 |
|  50 |     50 | GET_allreports.php           |           1 |    3 |
+-----+--------+------------------------------+-------------+------+
```

Figure 6 : MySQL table with request, session flag and role

Model set2 refers the list of web pages accessible by each role including web page sequence. Separate xml file is created for each role. Each tag in xml file other than a root element represents a page and list of possible accessible pages from that page. Figure 7 shows the 2 different xml sequence files for 2 different roles.

From the Figure 7, role1 user can access analysis.php, report.php, view.php and search.php pages from home.php. For role2 management.php, report.php, view.php and search.php are accessible from the home.php.

```
<?xml version='1.0' encoding="UTF-8">
<Pages>
<home.php>analysis.php, report.php, view.php,
search.php</home.php>
<analysis.php>declaration.php,publish.php
</analysis.php>
<view.php>viewusers.php,viewroles.php
</view.php>
</Pages>
```

```
<?xml version='1.0' encoding="UTF-8">
<Pages>
<home.php>management.php, report.php,
view.php, search.php</home.php>
<management.php>view.php,addusers.php
</management.php>
<view.php>viewusers.php,viewroles.php
</view.php>
</Pages>
```

Figure 7: Example of Sequence of pages accessed by role1 and role2

### 4.2.2. Runtime Enforcement Phase

During the enforcement phase, model is enforced and it continuously processes the web requests and web responses. It has **Enforcement Engine** and **Verifier Engine**. Figure 6 shows PROP Dynamic Analyzer functioning at Runtime Enforcement Phase.

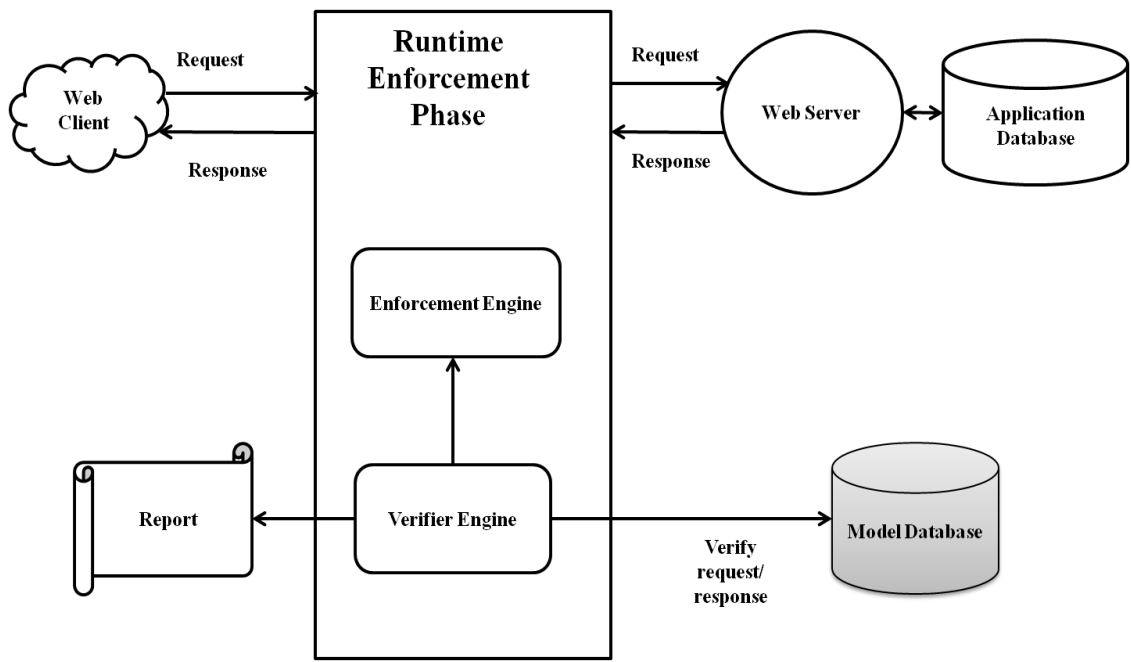

Figure 8 : PROP Dynamic Analyzer functioning at Runtime Enforcement Phase

**Enforcement Engine (EE)** captures the web request before hitting the web server directly. Captured request header information is sent to Verifier Engine. **Verifier Engine (VE)** checks the given request against model sets and sends the status of verification to the Enforcement Engine. If the request is a genuine behaviour of the application then the status represented as "don't_block" otherwise if any deviations occur with respect to model sets then represents status as "block" and logs an error based on diversion. Depending on the verification status Enforcement Engine proceeds further. If status is "don't_block" Enforcement Engine forwards the request to the web server application else it won't send the request to the web server application and intimate the web client about the diversion.

Verifier Engine first validates the web request information against Model Set1 which helps in finding authentication bypass attacks which generally happens in any vulnerable web application by changing the web page user input or by hijacking the session. Once the values satisfy the Model Set1 behaviour then it should go for next level of validation with respect to Model Set2 otherwise VE sends the "block" status to the EE which stops the web communication.

Once the authentication is done, the authorization check and sequence is verified with respect to Model Set 2. This verification addresses the vertical privilege escalation attacks where one role user is trying to access the pages of other roles and also addresses the sequence bypass attacks where the attacker is forcibly accessing the pages without following the sequence.

Once the request is satisfied with 2 levels of verification then only VE sends the "dont_block" status to Enforcement Engine, from there the request is passd to the web server. Failure at any level in the verification process, leads to VE sending "block" status to the EE and logging the error. EE reject the request and sends the error page to the user. The same process is followed for all requests.

## 5. IMPLEMENTATION DETAILS

### 5.1. Static Analyzer

PROP Static Analyzer is implemented with partial integration of open-source source code analysis tools Pixy, RIPS and PhpSecinfo. It targets PHP Configuration, XSS, SQL Injection, File Manipulation, File Inclusion, Command Injection and Code Injection related vulnerabilities in PHP based web applications. Also, it checks these vulnerabilities against user input, file and database related functions. After the complete scan of the web application it generates a well formed pdf report which will be useful for further analysis. The pdf report contains details about the scanned application name, time and date of the scan process and detected vulnerabilities information.

### 5.2. Dynamic Analyzer

#### 5.2.1. Training Phase

This phase generates the profile records by analyzing the request header information of each request. Request() captures the request and creates corresponding request profile files.

Collected profile records are analyzed and model database is created for that web application. Model database is represented in the form MySQL table and xml files. MySQL table contains request information of each page with the combination of communication id, request id, session flag and role. And separate xml files are created for each role. The xml file contains the sequence of pages accessed by each role.

**5.2.2. Runtime Enforcement Phase**

To analyze the application web requests coming from different web clients an apache module has been integrated with Runtime Enforcement Phase component of work flow analyzer. Apache module intercepts the request and sends the request to Verifier Engine for checking the requests against model database. Request is genuine, apache module forwards the same request to the web server otherwise blocks the request. IPC mechanism has been implemented between the Apache module and Verifier Engine.

We are considering the user-agent, client IP of each web request for differentiating the web clients and verifying against Model databases.

## 6. EXPERIMENTATION DETAILS

### 6.1. Static Analyzer

We have tested many applications with Static Analyzer and the details are mentioned below Table 1

Table 1: Tested Applications with Static Analyzer

| Application Name | Detected Vulnerabilities |
|---|---|
| Portal | SQL Injection<br>File Manipulation<br>Cross-site Scripting |
| Scarf | File Manipulation<br>SQL Injection<br>Cross-site Scripting |
| CET Automation tool | SQL Injection<br>Cross-site scripting |
| Bookstore | SQL Injection<br>Cross-site scripting |
| Employee_dir | SQL Injection<br>Cross-site scripting<br>File Manipulation |

### 6.2. Dynamic Analyzer

We have taken a web application with 2 roles manager and employer. Each user is having access to different web pages depending on the role. Table 2 shows the requests being made to the web application are represented in MySQL table during Training Phase.

Table 2: Model Database

| S.No | Communication id | Request id | Session | Role |
|---|---|---|---|---|
| 0 | 1 | GET_About.php | 0 | 0 |
| 1 | 2 | GET_Help.php | 0 | 0 |
| 2 | 3 | GET_Login.php | 0 | 0 |
| 3 | 4 | POST_Login.php | 0 | 0 |
| 4 | 5 | GET_Services.php | 0 | 0 |
| 5 | 6 | GET_Products.php | 0 | 0 |
| 6 | 7 | GET_home.php | 1 | manager |
| 7 | 8 | GET_Assign_works.php | 1 | manager |
| 8 | 9 | GET_User_mgmt.php | 1 | manager |
| 9 | 10 | GET_Update_users.php | 1 | manager |
| 10 | 11 | GET_Update_roles.php | 1 | manager |
| 11 | 12 | GET_View.php | 1 | manager |
| 12 | 13 | GET_Viewusers.php | 1 | manager |
| 13 | 14 | GET_Viewroles.php | 1 | manager |
| 14 | 15 | GET_Home.php | 1 | employer |
| 15 | 16 | GET_work_report.php | 1 | employer |
| 16 | 17 | GET_View.php | 1 | employer |
| 17 | 18 | GET_Viewusers.php | 1 | employer |
| 18 | 19 | GET_Viewroles.php | 1 | employer |

Table 3 & 4 shows the list of pages accessed and sequence of pages follows the current page by manager and employer respectively. For manager, possible list of pages accessible are Home.php, Assign_works.php, User_mgmt.php, View.php, Update_users.php, Update_roles.php, Viewusers.php and Viewroles.php. Viewusers.php & Viewroles.php can only accessible from View.php page. Means the user can not directly access the Viewusers.php from any of the page other than View.php.

Table 3: Role- manager

| Current Page | Next accessible pages |
|---|---|
| Home.php | Assign_works.php; User_mgmt.php; View.php |
| User_mgmt.php | Update_users.php; Update_roles.php |
| View.php | Viewusers.php; Viewroles.php |

Table 4: Role-employer

| Current Page | Next accessible pages |
|---|---|
| Home.php | Work_report.php; View.php |
| View.php | Viewusers.php; Viewroles.php |

According to Table 4, employer can access Home.php, Work_report.php, View.php, Viewusers.php and Viewroles.php pages.

## 7. PERFORMANCE DETAILS

### 7.1. Dynamic Analyzer

We deployed our security solution for PHP based web application and analyzed the performance. We tested the solution with observed page load time for different web pages by using lori add-on [25] installed in firefox with PROP Dynamic Analyzer Runtime Enforcement Phase and without PROP Dynamic Analyzer Runtime Enforcement Phase. With PROP, the curve is going slightly higher than without PROP because the runtime enforcement phase verifies each and every request against model sets and forwards the request to the web server. Figure 9 shows performance overhead with and without PROP.

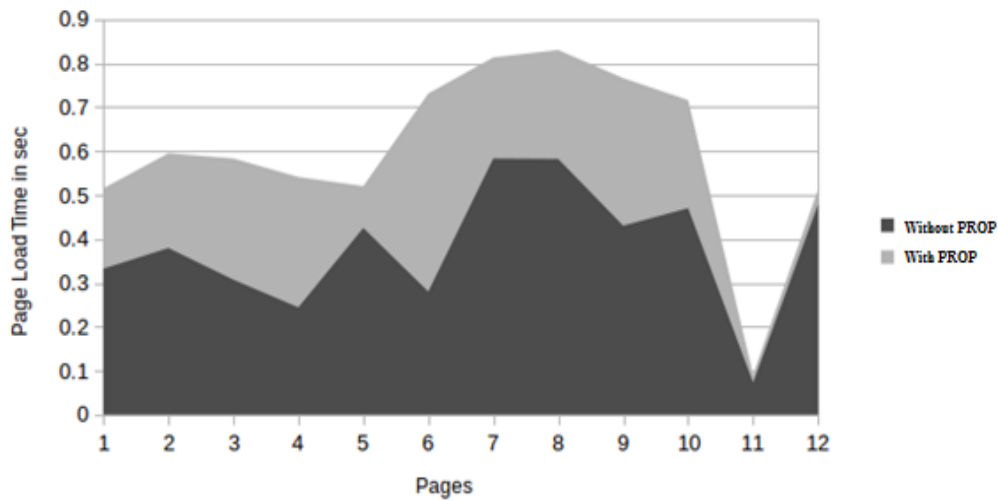

Figure 9 : Performance Overhead

## 8. CONCLUSIONS

In this paper, we discussed about the PHP source code analysis through Static Analyzer, monitoring web application behaviour at run time and an approach for detecting and preventing workflow deviations through Dynamic Analyzer. PROP security solution identifies PHP source code vulnerabilities like XSS, SQL Injection, Code Injection, Command Injection, File Inclusion, File Manipulation and PHP configuration vulnerabilities with Static Analyzer component and authentication bypass, session hijacking and sequence bypass attacks with Dynamic Analyzer component. A pdf report has been generated to analyse the vulnerabilities details in case of Static Analyzer and deviation logs are maintained in case of Dynamic Analyzer.

Furthermore, it addresses the authorization bypass attack if users with role binding details are known in the prior. PROP Dynamic Analyzer has its own limitation like, it is able to detect and prevent Authorization bypass for different role users, but users within the same role is trying to bypass is not addressed. By checking the user's session at every page can address this issue. Another limitation is we are crawling the site by considering the href links and opening the authentication pages (login and logout pages) for different roles in a browser which may not cover all the web pages. If the web site is designed with form based actions where the manual interaction is mandatory to pass the parameters, automatic crawl may not encompass. To overcome this problem we are attempting to render the form action based pages in browser to collect the data from the user which aids to crawl to next page.

## ACKNOWLEDGEMENTS

Our sincere thanks to Department of Electronics & Information Technology (Deity), Ministry of Communications and Information Technology, Government of India for supporting this research work.

## Authors

Mrs Sireesha Chiliveri is presently working at Centre for Development of Advanced Computing(C-DAC), Hyderabad as a Technical officer. She is associated with C-DAC from last 8 years in R&D of System Software and Network Security domain. Her areas of interest include Web Application Security, End System Security, Malware Analysis and Linux System Programming.

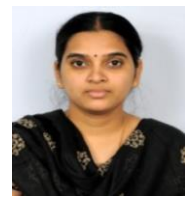

Mrs Jyostna G is presently working as a Senior Technical Officer at Centre for Development of Advanced Computing (C-DAC), Hyderabad. She is associated with C-DAC from last 9 years in R&D of System Software and Network Security domain. Her areas of interest include Web Application Security, End System Security, Malware Analysis and Linux System Programming. She has around 9 publications in International Journals/conferences.

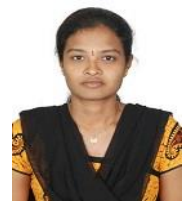

Mr Raghu Varan Reddy P is associated with R&D, Centre for Development of Advanced Computing(C-DAC), Hyderabad from last 3 years in System Software and Network Security domain. His areas of interest include Web Application Security, End System Security and Linux System Programming & Driver Development.

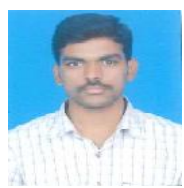

Mrs P.R.Lakshmi Eswari, is presently working as Joint Director, e-Security R&D, C-DAC Hyderabad. She is currently involved in the Research & Development of end system security solutions focusing on anti Malware & device control solutions. As an outcome of R&D efforts solutions like USB Pratirodh, AppSamvid, Browser JS Guard, Malware Resist etc. are developed by their team. She is associated with C-DAC for the past 14 years in various R&D projects and training activities. Earlier she worked as lecturer at NIT Warangal for 2 years. She did her B.E and M.Tech in Computer Science and Engg stream and currently pursuing her PhD with JNTU. She has around 12 publications in International Journals/conferences.

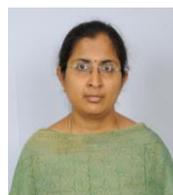